\documentclass[onecolumn,draftclsnofoot]{IEEEtran}

\usepackage{amsmath}
\usepackage{theorem}
\usepackage{epsfig}
\usepackage{amsfonts}
\usepackage{amssymb}
\usepackage{mathrsfs}
\usepackage{euscript}

\usepackage{graphicx}
\usepackage{multirow}
\renewcommand{\baselinestretch}{2}

\title{On The Achievable Rate Region of a New Wiretap Channel With Side Information}

\author{$^{\dag}$ Hamid G. Bafghi, $^{\dag}$Babak Seyfe, $^{\ddag}$Mahtab Mirmohseni,$^{\ddag}$M. Reza Aref\\
\small {$^{\dag}$ Electrical Engineering Department, Shahed University, Tehran, Iran.}\\
$^{\dag\ddag}$ ISSL Laboratory, Electrical Engineering Department, Sharif University of Technology, Tehran, Iran.\\
Emails: \{ghanizade, seyfe\}@shahed.ac.ir,
mirmohseni@ee.sharif.edu, aref@sharif.edu
}

\begin{document}
\maketitle
\pagenumbering{arabic}
\pagestyle{empty}
\pagestyle{plain}

\begin{abstract}

~A new applicable wiretap channel with separated side information is considered here which consist of a sender, a legitimate receiver and
a wiretapper. In the considered scenario, the links from the transmitter to the legitimate receiver and the eavesdropper experience different conditions or channel states. So, the legitimate receiver and the wiretapper listen to the transmitted signal through the channels with different channel states which may have some correlation to each other. It is assumed that the transmitter knows the state of the main channel non-causally and uses this knowledge to encode its message. The state of the wiretap channel is not known anywhere. An achievable equivocation rate region is derived for this model and is compared to the existing works. In some special cases, the results are extended to the Gaussian wiretap channel.
\\
\end{abstract}

\section{Introduction}

Secure communication from an information theoretic perspective was first studied by Shannon in his famous paper~\cite{bibi28}, where a noiseless channel model was assumed with an eavesdropper which has an identical copy of the encrypted message as a legitimate receiver, and the sufficient and necessary condition for perfect secrecy using information theoretic concepts were established. In the Shannon's model, a source message~$W$ is encrypted to a ciphertext~$E$ by a key~$K$ shared by the transmitter and the receiver. An eavesdropper, which knows the family of encryption functions, i.e., keys and the probability of choosing the keys, may intercept the ciphertext~$E$. The system is considered to be perfectly secure if the a posteriori probabilities of~$W$ for all~$E$ would be equal to the a priori probabilities, i.e.,~$P(W|E)= P(W)$. Alternatively, Shannon proved that the perfect secrecy can be achieved only when the secret key is at least as long as the plaintext message or more precisely, when~$H(K)\geq H(W)$.

\begin{figure}
\centering
\epsfig{file=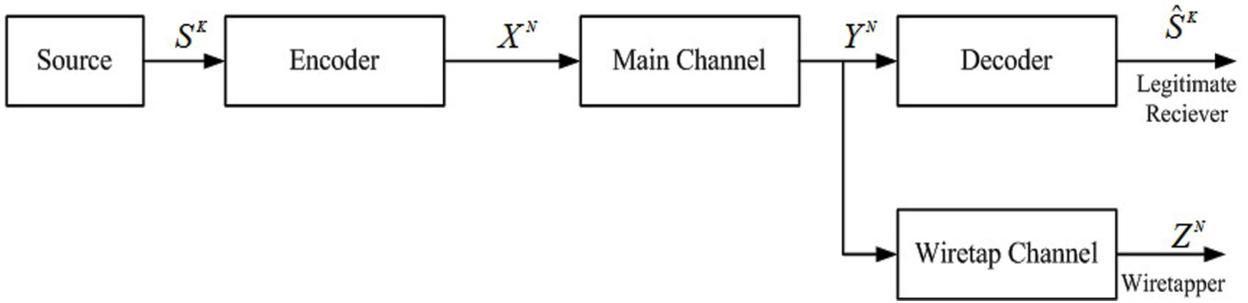,width=1\linewidth,clip=}
\caption{Wyner's Wiretap channel~\cite{bibi7}.
In this channel, it is assumed that the channel to the eavesdropper
is physically degraded version of the channel to the legitimate receiver.}
\label{fig:1}
\end{figure}

The wiretap channel was first introduced and studied by Wyner in his fundamental paper~\cite{bibi7} which is the most basic physical layer model  explains the communication security's problems. In his model, the transmitter wishes to transmit a source signal, i.e., a confidential message, to a legitimate receiver in a way that this message be kept secret from an eavesdropper. In this model illustrated in Fig.~\ref{fig:1}, despite of the Shannon's model, it is assumed that the channel to the eavesdropper is a physically degraded version of the channel to the legitimate receiver. In other words, the channel's output at the eavesdropper may be a noisy version of the channel output at the legitimate receiver. On the other hand, the transmitter communicates to the intended receiver through the main channel which may be noisy or noiseless, but the wiretapper receives a noisy copy of the message through a wiretap channel which is a cascade of the main channel. In addition, Wyner~\cite{bibi7} assumed that the eavesdropper knows the transmitter's encoding-decoding scheme. So, the objective is maximizing the rate of reliable communication such that the wiretapper realizes as little as possible about the source output. The information leakage was measured by equivocation rate as~$\Delta\triangleq H(S^{K}|Z^{N})$, where~$S^{K}$ and~$Z^{N}$ are represented the message set and the channel output at the wiretapper, respectively.
Eavesdropper is assumed to be a passive receiver which does not transmit any signal over the channel. Furthermore, Wyner~\cite{bibi7} proposed a basic principle coding strategy to achieve secure communication for wiretap channels which is based on the fact that the eavesdropper is not able to decode any information more than it's channel capacity.

Csisz\'{a}r and K\"{o}rner generalized the Wyner's wiretap channel~\cite{bibi8}. In their model, it is assumed that the wiretap channel's output is not necessarily a degraded version of the legitimate receiver's one. They showed that the secrecy capacity can be expressed as~$C_{s}=max_{U\rightarrow X\rightarrow (Y,Z)}[I(U;Y)-I(U;Z)]$, where~$X$,~$Y$ and~$Z$ are the channel input, the channel output in the legitimate receiver and the channel output at the wiretapper, respectively. Moreover, the maximization is over all random variables~$U$ in joint distribution with~$X$,~$Y$ and~$Z$ such that~$U\rightarrow X\rightarrow (Y,Z)$ forms a Markov Chain.

Using the channel state information in communication channel models was introduced by Shannon in his landmark paper~\cite{bibi39}, where he assumed the availability of Channel Side Information at the Transmitter (CSIT). Gel'fand and Pinsker in their essential work~\cite{bibi22} proved that the capacity of the state-dependent discrete memoryless channel with non-causally CSIT is given by~$C=max_{p(u,x|v)}[I(U;Y)-I(U;V)]$, where the maximum is taken over all input distribution~$p(u,x|v)$ with a finite alphabet auxiliary random variable~$U$.

Costa in his well known paper named \emph{Writing on Dirty Paper}, extended this result to the Gaussian channel and showed that for this channel, interference did not affect the capacity~\cite{bibi30}. He chose~$U= X+ \alpha V$ and maximized the Gel'fand and Pinsker's capacity over all quantity of~$\alpha$ and proved that for this value of~$\alpha$, the capacity of the channel reduces to the channel without states. The dirty paper channel was extended to the basic Gaussian wiretap channel with side information by Mitrpant and et al.~\cite{bibi24}, in which an achievable and upper bound for this channel has been introduced.

\begin{figure}
\centering
\epsfig{file=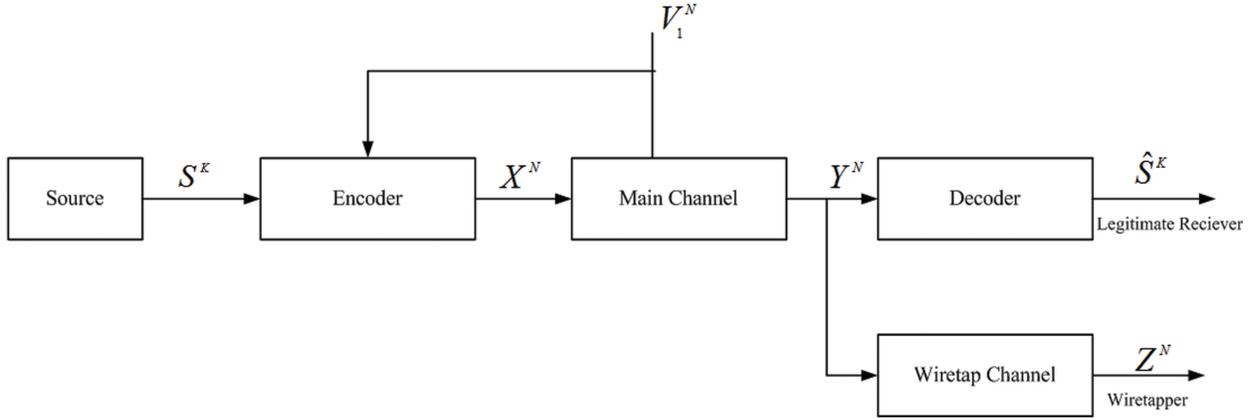,width=1\linewidth,clip=}
\caption{Wiretap channel with side information introduced by Chen and Vinck~\cite{bibi23}.}
\label{fig:2}
\end{figure}

Chen and Vinck investigated Wyner's wiretap channel with side information~\cite{bibi23} (Fig.~\ref{fig:2}). Their results are based on the previous wiretap channel's results in~\cite{bibi7},~\cite{bibi8},~\cite{bibi24} and the discrete memoryless channel with state information~\cite{bibi22}. They gave an achievable rate region which is established using a combination of the Gel'fand–-Pinsker coding and the Wyner's wiretap coding. They extended their results to the Gaussian wiretap channel with side information using the same technique like dirty paper channel~\cite{bibi23}.

\begin{figure}
\centering
\epsfig{file=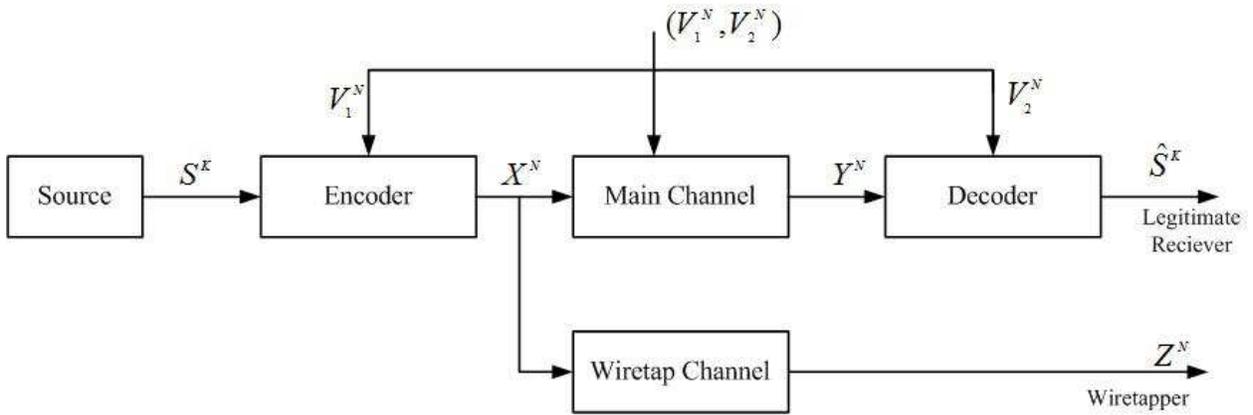,width=1\linewidth,clip=}
\caption{Wiretap channel with two--sided channel state information~\cite{bibi41}.}
\label{fig:3}
\end{figure}
Furthermore, there were some different works on the wiretap channel with and without side information. The work~\cite{bibi43} studied the two way wiretap channel. The Gaussian wiretap channel with m-pam inputs was considered in~\cite{bibi44} and the secrecy capacity of the Gaussian MIMO multi-receiver wiretap channel was investigated by~\cite{bibi45}. Liu et al. in~\cite{bibi41}, studied the two-sided channel state problem in the discrete memoryless wiretap channel, where as shown in Fig.~\ref{fig:3}, the information of the two-sided channel states are available at the transmitter and the main receiver, respectively. In addition, in their scenario the wiretap channel is not necessarily a degraded version of the main channel. An achievable rate equivocation region for this general case is given in~\cite{bibi41}. Khisti et. al., considered the secret-key agreement problem in the wiretap channel~\cite{bibi40},~\cite{bibi29}. In their model, the transmitter communicates to the legitimate receiver and the eavesdropper over a discrete memoryless wiretap channel with a memoryless state sequence. The transmitter and the legitimate receiver generate a shared secret key that remains secret from the eavesdropper. The results are comparable to the wiretap channel introduced by~\cite{bibi23}. Recently, an improved lower bound for the wiretap channel with causal state information at the transmitter and receiver has been reported in~\cite{bibi42}, where the achievability of the rate region is proved using block Markov coding, Shannon strategy, and key generation from the common state information~\cite{bibi39}. The state sequence available at the end of each block, is used to generate a key which is used to enhance the transmission rate of the confidential message in the following block.

\begin{figure}
\centering
\epsfig{file=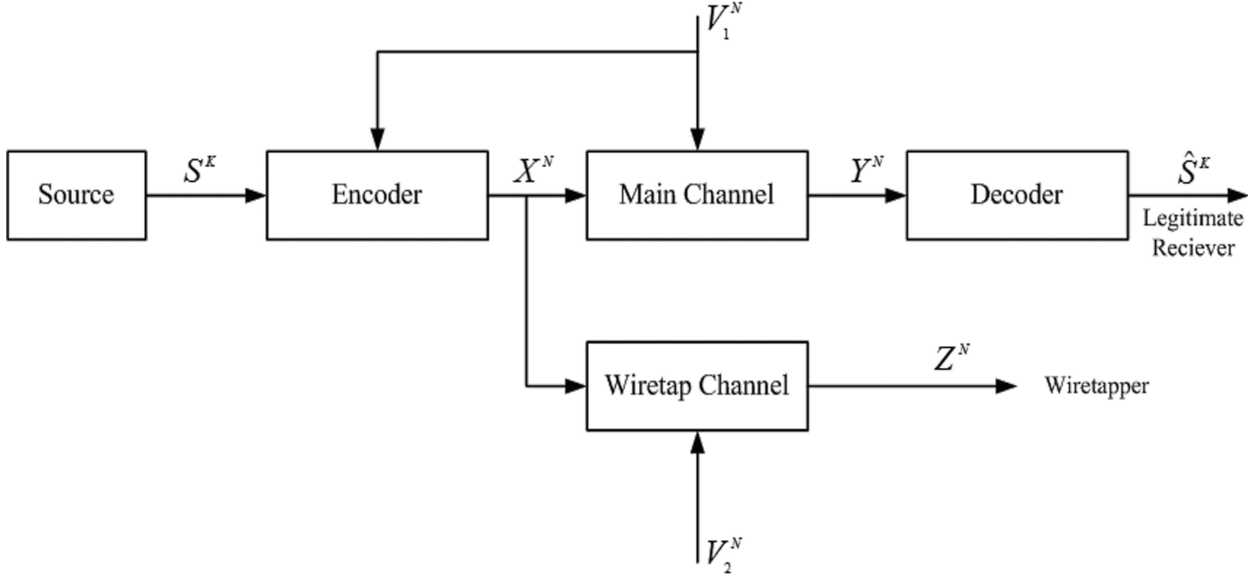,width=1\linewidth,clip=}
\caption{The new more general wiretap channel with side information. The wiretapper's messages is not a degraded version of the legitimate receiver's one and the receiver and the wiretapper listen to the sent message from the separated channels with different channel states.}
\label{fig:4}
\end{figure}

In this paper, we introduce a new wiretap channel model with side information, in which the wiretapper's messages is not a degraded version of the legitimate receiver's one. On the other hand, the transmitter sends its message through the main and the wiretap channels. So, the receiver and the wiretapper listen to the sent message from the separated channels with different characteristics, i.e., different channel states. This model is a general case of Chen--Vinck~\cite{bibi23} and Wyner wiretap channel~\cite{bibi7} and reduces to these channels in special cases. We extend our model to the Gaussian wiretap channel where the states of the main and wiretapper channels are different with some correlation coefficients. In the Gaussian case, if the correlation coefficients are equal to one, our channel reduces to Chen--Vinck's channel. The proposed channel is illustrated in Fig.~\ref{fig:4}.
The rest of the paper is organized as follows. In Section~\ref{II}, the channel model is introduced. The main results are presented in Section~\ref{III}.
In Section~\ref{IV}, the proof of the main results are given. In Section~\ref{V}, the results are extended to the Gaussian case and the paper is
concluded in the last section.
\section{Channel Model and Preliminaries}\label{II}
First, we clear our notation in this paper. Let~$\mathcal{X}$ be a finite set. Denote its cardinality by~$|\mathcal{X}|$. If we consider~$\mathcal{X}^{N}$, the members of~$\mathcal{X}^{N}$ will be written as~$x^{N} = (x_{1}, x_{2}, ..., x_{N})$, where subscripted
letters denote the components and superscripted letters denote the vector. A similar convention applies to random vectors and random
variables, which are denoted by uppercase letters.

Consider the situation shown in Fig.~\ref{fig:4}. Assume that the state information of the main channel, i.e., the channel from the transmitter to the legitimate receiver, is known at the encoder non-causally but the state of the wiretapper's channel is unknown and the channels' states, i.e. ~$V_{ti}$,~$t=1, 2$,~$1\leq i\leq N$, are independent and identically distributed (i.i.d), but~$V_{1i}$ and~$V_{2i}$ are correlated. The transmitter sends the message~$s^{k}\in\{1,2,\ldots,M\}$ to the legitimate receiver in~$N$ channel uses. Based on the~$s^{k}$ and~$v^{N}$, the encoder generates the codeword~$x^{N}$ and transmits it on the main and the wiretap channels. The decoder at the legitimate receiver makes an estimation of the transmitted message~$\hat{s}^{k}$ based on the received message~$y^{N}$. The corresponding output at the wiretapper is~$z^{N}$. The channels are memoryless, i.e.,
\begin{eqnarray}\label{eqn1}
p(y^{N}|x^{N}, v^{N})=\prod^{N}_{i=1}p(y_{i}|x_{i}, v_{i})\label{eqn2}\\
p(z^{N}|x^{N}, v^{N})=\prod^{N}_{i=1}p(z_{i}|x_{i}, v_{i})
\end{eqnarray}
Assume that~$S^{k}$ is uniformly distributed on~$\{1,2,\ldots, M\}$, so~$H(S^{k})=\log M$. The average probability of error~$P_{e}$ is given by
\begin{eqnarray}\label{eqn3}
P_{e}=\frac{1}{M}\sum^{M}_{i=1}Pr(\hat{S}^{k}(Y^{N})\neq i|S^{k}=i)
\end{eqnarray}
We define the rate of the transmission to the intended receiver to be
\begin{eqnarray}\label{eqn4}
R=\frac{\log M}{N}
\end{eqnarray}
and the fractional equivocation wiretapper to be
\begin{eqnarray}\label{eqn5}
d=\frac{H(S^{k}|Z^{N})}{H(S^{k})}
\end{eqnarray}

Obviously, we have~$H(S^{k}|Z^{N})=NRd$.

\section{MAIN RESULTS: \\outer and inner bounds}\label{III}

Like~\cite{bibi23}, we say that the pair~$(R^{*},d^{*})$ is achievable, if for all~$\epsilon >0$, there exists an encoder-decoder pair such that
\begin{eqnarray}\label{eqn6}
R\geq R^{*}-\epsilon, d\geq d^{*}-\epsilon, P_{e}\leq \epsilon.
\end{eqnarray}

\emph{Definition 1:} The \textit{secrecy capacity}~$C_{s}$ is the maximum~$R^{*}$ such that~$(R^{*},1)$ is achievable.

\emph{Definition 2:} We denote

\begin{eqnarray}\label{eqn7}
\lefteqn{R_{U1}=I(U;Y)-\max \{I(U;V_{1},V_{2}), I(U;Z)\}}\\\label{eqn8}
&R_{U2}=I(U;Y)- I(U;V_{1},V_{2})\\\label{eqn9}
&d_{U2}=\frac{R_{U1}}{R_{U2}}=\frac{I(U;Y)-\max \{I(U;V_{1},V_{2}), I(U;Z)\}}{I(U;Y)- I(U;V_{1},V_{2})}
\end{eqnarray}
where~$U$ is an auxiliary random variable such that~$U\rightarrow (X,V_{1},V_{2})\rightarrow (Y, Z)$
forms a Markov chain. Now, consider the following result:

\emph{Theorem 1:} For the discrete memoryless channel with side information shown in Fig.~\ref{fig:4}, we denote~$\mathcal{R}_{U}$ as the
set of points~$(R, d)$ with~$R_{U1}\leq R\leq R_{U2}$,~$0\leq d\leq 1$ and~$Rd=R_{U1}$. Let
\begin{eqnarray}\label{eqn10}
\mathcal{R}^{'}_{U}\triangleq \{(R^{'}, d^{'}): 0\leq R^{'}\leq R, 0\leq d^{'}\leq d, (R, d)\in \mathcal{R}_{U}\}.
\end{eqnarray}

Then the set~$\mathcal{R}$, defined as following, is achievable:
\begin{eqnarray}\label{eqn11}
\mathcal{R}=\bigcup_{U\rightarrow (X,V_{1},V_{2})\rightarrow (Y, Z)} \mathcal{R}^{'}_{U}.
\end{eqnarray}

The region is achievable if we limit the cardinality of~$U$ by the constraint~$|\mathcal{U}|\leq |\mathcal{X}||\mathcal{V}_{1}||\mathcal{V}_{2}|+4$.

\begin{proof}
The proof of the theorem is relegated to the next Section. The constraint is implied by lemma 3 of~\cite{bibi25}.
\end{proof}
\emph{Remark 1:} The point~$(R, d)$ in~$\mathcal{R}$ with ~$d=1$ is of considerable interest. These situations correspond to the perfect secrecy situation, defined as
\begin{eqnarray}\label{eqn12}
R_{s}=\max_{U\rightarrow (X, V_{1}, V_{2})\rightarrow (Y, Z)}R_{U1}
\end{eqnarray}

The following theorem bounds the secrecy capacity of the proposed wiretap channel with the side information.

\emph{Theorem 2:} For the discrete memoryless wiretap channel with side information, shown in Fig.~\ref{fig:4}, we have
\begin{eqnarray}\label{eqn13}
R_{s}\leq C_{s}\leq \min \{C_{M}, \max_{{U\rightarrow (X,V_{1},V_{2})\rightarrow (Y, Z)}} [I(U;Y)-I(U;Z)]\}
\end{eqnarray}
where~$C_{M}$ is the capacity of the main channel.

\begin{proof} From Theorem 1, we have~$R_{s}\leq C_{s}\leq C_{M}$ and from the result by Csisz\'{a}r and K\"{o}rner~\cite{bibi8} we have
 $C_{s}\leq \max_{U \rightarrow (X,V_{1},V_{2})\rightarrow (Y, Z)} [I(U;Y)-I(U;Z)]$.
This completes the proof.
\end{proof}
\section{The Proof of Theorem 1}\label{IV}

In this Section, we prove the achievability of the region~$\mathcal{R}$. We prove that the rate equivocation pairs~$(R_{U1}, 1)$ and~$(R_{U2}, d_{U2})$ are achievable and then by implying time--sharing, achievability of the region~$\mathcal{R}^{'}_{U}$ is proved.

\subsection{$(R_{U1}$, 1) is Achievable}\label{IV-A}

First we construct random codebooks by the following generation steps:

\subsubsection{Codebook Generation}.

\textbf{a.} Generate~$2^{N[I(U;Y)-\epsilon_{UY}]}$ i.i.d sequences~$u^{N}$, according to the distribution~$p(u^{N})=\prod ^{N}_{i=1}p(u_{i})$.

\textbf{b.} Partition these~$u^{N}$ sequences into~$2^{NR}$ bins where $R=[R_{U1}-\epsilon _{UY}-\epsilon_{UV_{1}V_{2}Z}]$. Index each bin by~$j\in \{1, 2, \ldots, 2^{NR}\}$. Thus each bin contains $2^{N[max\{I(U;V_{1},V_{2}),I(U;Z)\}+\epsilon _{UV_{1}V_{2}Z}]}$ sequences.

\textbf{c.} Distribute $2^{N[max\{I(U;V_{1},V_{2}),I(U;Z)\}+\epsilon _{UV_{1}V_{2}Z}]}$ sequences randomly into $2^{N[max\{ I(U;V_{1},V_{2}),I(U;Z)\}- I(U;Z)+\epsilon _{UV_{1}V_{2}Z}+\epsilon_{UZ}]}$ subbin such that every subbin contains $2^{N[I(U;Z)-\epsilon_{UZ}]}$ sequences. Then index each subbin which contains~$U^{N}$ by
\begin{eqnarray}
w\in \{1, 2, \ldots 2^{N[max\{I(U;V_{1},V_{2}),I(U;Z)\}-I(U;Z)+\epsilon _{UV_{1}V_{2}Z}+\epsilon_{UZ}]}\}\nonumber.
\end{eqnarray}

\subsubsection{Encoding}
To transmit message~$j$ thorough the main channel with interference~$v_{1}^{N}$, the transmitter finds~$j$-th bin of the sequence~$u^{N}(j)$ such that~$(u^{N}, v_{1}^{N})\in T^{N}_{\epsilon}(P_{UV_{1}})$. We use~$T^{N}_{\epsilon}(
P_{UV_{1}})$ to denote the strong typical set based on the distribution~$P_{UV_{1}}$, otherwise choose~$j=1$. The transmitter sends the
associated jointly typical~$x^{N}(j)$ generated according to ~$p(x^{N}(j) | u^{N}(j), v_{1}^{N}) = \prod_{i=1}^{N} p(x_{i}|u_{i}, v_{1,i})$

\subsubsection{Decoding}
The intended receiver receives~$y^n$ according to the distribution~$\prod_{i=1}^{N} p(y_{i}|x_{i},v_{1,i})$. Then it looks for the unique sequence~$u^{N}$ such that~$(u^{N},v_{1}^{N})\in T^{N}_{\epsilon}(P_{UV_{1}})$ and the index of the bin containing~$u^{N}$ is declared as the transmitted message.

\subsubsection{Wiretapper}
The wiretapper receives a sequence~$z^{N}$ according to~$\prod^{N}_{i=1} p(z_{i}|x_{i},v_{2,i})$.

Now, we prove that~$(R_{U1},1)$ is achievable. As the first step we should prove that~$P_{e}\longrightarrow 0$, as~$N\rightarrow \infty$. Our
encoding-decoding strategy is similar to the one used in~\cite{bibi23} and it is easy to show that the information rate~$R_{U1}$ in the main channel
is achievable. For more detail see Appendix A in~\cite{bibi23}.
As the second step, we should prove that~$d\rightarrow 1$, as~$N\rightarrow\infty$. In this step, we consider the uncertainty of the message to the wiretapper. So we have

\begin{eqnarray}\label{eqn14}
\lefteqn{H(S^{k}| Z^{N})}\nonumber\\
&\!\!\!\!\!\!\!\!\!\!\!\!\!\!\!\!\!\!\!\!\!\!\!\!\!\!\!\!\!\!\!\!\!\!\!\!\!\!\!\!\!\!\!\!\!\!\!\!\!\!\!\!\!\!\!\!\!\!\!\!\!\!\!\!\!\!
\!\!\!\!\!\!\!\!\!\!\!\!\!\!\!\!\!\!\!\!\!\!\!\!\!\!\!\!\!\!\!\!\!\!\!\!\!\!\!\!\!\!\!\!\!\!\!\!\!\!\!\!\!\!\!\!\!\!\!\!\!\!\!\!\!\!
\!\!\!\!\!\!\!\!\!\!\!\!=H(S^{k}, Z^{N})- H(Z^{N})\nonumber\\
&\!\!\!\!\!\!\!\!\!\!\!\!\!\!\!\!\!\!\!\!\!\!\!\!\!\!\!\!\!\!\!\!\!\!\!\!\!\!\!\!\!\!\!\!\!\!\!\!\!\!\!\!\!\!\!\!\!\!\!\!\!\!\!\!\!\!
\!\!\!\!\!\!\!\!\!\!\!\!\!\!\!\!\!\!\!\!\!\!\!\!= H(S^{k}, Z^{N}, W)- H(W|S^{k},Z^{N})- H(Z^{N})\nonumber\\
&\!\!\!\!\!\!\!\!\!\!\!\!\!\!\!\!\!\!\!\!\!\!\!= H(S^{k}, Z^{N}, W, U^{N})- H(U^{N}| S^{k}, Z^{N}, W)- H(W| S^{k}, Z^{N})- H(Z^{N})\nonumber\\
& \quad\quad\quad\!\!= H(S^{k},  W| Z^{N}, U^{N})+ H(U^{N}, Z^{N}) - H(U^{N}| S^{k}, Z^{N}, W)- H(W| S^{k}, Z^{N})-H(Z^{N})\nonumber\\
&\!\!\!\!\!\!\!\!\!\!\!\!\!\!\!\!\!\!\!\!\!\!\!\!\!\!\!\!\!\!\!\!\!\!\!\!\!\!\!\!\!\!\!\!\!\!\!\!\!\!\!\!\!\!\!\!\!\!\!\!\!
\geq^{(a)} H(U^{N}| Z^{N}) - H(U^{N}| S^{k}, Z^{N}, W)- H(W| S^{k}, Z^{N})\nonumber\\
&\!\!\!\!\!\!\!\!\!\!\!\!\!\!\!\!\!\!\!\!\!\!\!\!\!\!\!\!\!\!\!\!\!\!\!\!\!\!\!\!\!
\geq^{(b)} H(U^{N}| Z^{N}) - H(U^{N}| S^{k}, Z^{N}, W)- \log|\mathcal{W}|- H(U^{N}| Y^{N})\nonumber\\
&\!\!\!\!\!\!\!\!\!\!\!\!\!\!\!\!\!\!\!\!\!\!\!\!\!\!\!\!\!\!\!\!\!\!\!\!\!\!\!\!\!\!\!\!\!\!\!\!\!\!\!\!\!\!\!\!\!\!\!\!\!\!\!\!\!
\!\!\!\!\!\!\!\!\!\!\!\!\!\! =^{(c)} N[I(U; Y)- I(U; Z)]- H(U^{N}| S^{k}, Z^{N}, W)\nonumber\\
&\!\!\!\!\!\!\!\!\!\!\!\!\!\!\!\!\!\!\!\!\!\!\!\!\!\!\!\!\!\!\!\!\!\!\!\!\!\!\!\!\!\!\!\!\!\!\!\! - N[max\{I(U;V_{1},V_{2}),I(U;Z)\}-I(U;Z)+\epsilon _{UV_{1}V_{2}Z}+\epsilon_{UZ}]\nonumber\nonumber\\
&\!\!\!\!\!\!\!\!\!\!\!\!\!\!\!\!\!\!\!\!\!\!\!\!\!\!\!\!\!\!\!\!\!\!\!\!\!\!\!\!\!\!\!\!\!\!\!\!\!\!\!\!\!\!\!\!\!\!\!\!\!\!\!\!\!\!\!\!\!\! =NR_{U1}- H(U^{N}| S^{k}, Z^{N}, W)- N[\epsilon _{UV_{1}V_{2}Z}+\epsilon_{UZ}]\nonumber\\
\end{eqnarray}
where

$(a)$ follows from the fact that~$H(S^{k},  W| Z^{N}, U^{N})\geq 0$;

$(b)$ is because of the fact that~$H(W| S^{k}, Z^{N})\leq H(W)\leq \log |\mathcal{W}|$ and~$H(U^{N}| Y^{N})\geq 0;$

$(c)$ follows from the fact that~$I(U^{N}; Y^{N})= NI(U; Y)$,~$I(U^{N}; Z^{N})= NI(U; Z)$ and
\begin{eqnarray}
log| \mathcal{W}| = N [max\{I(U;V_{1},V_{2}),I(U;Z)\}-I(U;Z)+\epsilon _{UV_{1}V_{2}Z}+\epsilon_{UZ}].\nonumber
\end{eqnarray}

To compute the second term in~(\ref{eqn14}), we should bound the entropy of the codeword conditioned on the bin~$j$, subbin~$w$ and the wiretapper's received signal~$z^{N}$. We consider the subbin~$w$ in bin~$j$ as a codebook,~$U^{N}$ in the codebook as the input message and~$Z^{N}$ as the result of passing~$U^{N}$ through the wiretap channel. From~$Z^{N}$, the decoder estimates the sent message~$U^{N}$. Let~$g(\cdot)$ be the decoder and the estimate be~$\hat{U}^{N}= g(\cdot)$. Define the probability of error

\begin{eqnarray}\label{eqn15}
P_{SB}= Pr\{\hat{U}^{N}\neq U^{N}\}.
\end{eqnarray}

By Fano's inequality~\cite{bibi34}, we have

\begin{eqnarray}\label{eqn16}
H(U^{N}| S^{k}=j, W= w, Z^{N})\leq h(P_{SB})+ P_{SB}N[I(U; Z)-\epsilon_{UZ}].
\end{eqnarray}

Hence

\begin{eqnarray}\label{eqn17}
H(U^{N}| S^{k}, W, Z^{N})\leq h(P_{SB})+ P_{SB}N[I(U; Z)-\epsilon_{UZ}].
\end{eqnarray}

Now, we should prove that for arbitrary~$0<\lambda< 1/2$,~$P_{SB}\leq \lambda$. The proof is similar to the one in~\cite{bibi23}. Thus, we have bounded~$P_{SB}$ for given arbitrary small~$\epsilon$ and~$\lambda$.

Combining~(\ref{eqn5}),~(\ref{eqn14}),~(\ref{eqn17}) and the bound on~$P_{SB}$ we have
\begin{eqnarray}\label{eqn18}
d\geq 1-\frac{\epsilon_{UZ}- \epsilon_{UY}- h(\lambda)/N+ \lambda[I(U; Z)- \epsilon _{UZ}]}{R_{U1}- \epsilon_{UY}- \epsilon_{UV_{1}V_{2}Z}}.
\end{eqnarray}

Thus we derive that~$d\rightarrow 1$, as~$N\rightarrow\infty$.
\subsection{$(R_{U2}, d_{U2})$ is Achievable}

From the~(\ref{eqn7})-~(\ref{eqn9}), it is derived that if~$I(U; V_{1}, V_{2})\geq I(U; Z)$, then the equivocation rate pair~$(R_{U2}, d_{U2})$ is equal with~$(R_{U1}, 1) $. So, we should prove that if~$I(U; V_{1}, V_{2})< I(U; Z)$, then~$(R_{U2}, d_{U2})$ is achievable. In this case, when~$I(U; V_{1}, V_{2})< I(U; Z)$, we have
\begin{eqnarray}\label{eqn19}
\lefteqn{R_{U2}= I(U; Y)- I(U; V_{1}, V_{2})}\label{eqn20}\\
&d_{U2}= \frac{I(U; Y)- I(U; Z)}{I(U; Y)- I(U; V_{1}, V_{2})}
\end{eqnarray}

Now we introduce the encoding and decoding strategy.

\subsubsection{Codebook Generation}.

\textbf{a.} Generate~$2^{N[I(U;Y)-\epsilon_{UY}]}$ i.i.d sequences~$u^{N}$, according to the distribution~$p(u^{N})=\prod ^{N}_{i=1}p(u^{N})$.

\textbf{b.} Partition these sequences into~$2^{NR}$ bins where~$R=[R_{U2}-\epsilon _{UY}-\epsilon_{UV_{1}V_{2}}]$. Index each bin by~$j\in \{1, 2, \ldots, 2^{NR}\}$. Thus each bin contains~$2^{N[I(U;V_{1},V_{2})+\epsilon _{UV_{1}V_{2}Z}]}$ sequences.

\textbf{c.} Distribute~$2^{N[max\{I(U;V_{1},V_{2}),I(U;Z)\}+\epsilon _{UV_{1}V_{2}Z}]}$ sequences randomly into~$2^{N[max\{ I(U;V_{1},V_{2}),I(U;Z)\}- I(U;Z)+\epsilon _{UV_{1}V_{2}Z}+\epsilon_{UZ}]}$ subbins such that every subbin contains~$2^{N[I(U;Z)-\epsilon_{UZ}]}$ sequences. Then index each subbin containing~$U^{N}$ by~$w\in \{1, 2, \ldots 2^{N[max\{I(U;V_{1},V_{2}),I(U;Z)\}-I(U;Z)+\epsilon _{UV_{1}V_{2}Z}+\epsilon_{UZ}]}\}$.

\subsubsection{Encoding}
To transmit message~$j$ thorough the main channel with interference~$v_{1}^{N}$, transmitter finds bin~$j$ for a sequence~$u^{N}(j)$ such that~$(u^{N}, v_{1}^{N})\in T^{N}_{\epsilon}(P_{UV_{1}})$, otherwise choose~$j=1$.

\subsubsection{Decoding}
The intended receiver receives~$y^n$ according to the distribution~$\prod_{i=1}^{N} p(y_{i}|x_{i},v_{1,i})$. Then the receiver looks for the unique sequence~$u^{N}$ such that~$(x^{N},v_{1}^{N})\in T^{N}_{\epsilon}(P_{UV_{1}})$ and the index bin of the bin containing~$u^{N}$ declares as the message index.

\subsubsection{Wiretapper}
The wiretapper receives a sequence~$z^{N}$ according to~$\prod^{N}_{i=1} p(z_{i}|x_{i},v_{2,i})$.

To prove that~$(R_{U2}, d_{U2})$ is achievable, first we should prove that~$P_{e} \rightarrow 0$, as~$N \rightarrow \infty$. The proof is similar to the one in Section~\ref{IV-A}. Then we should prove that~$d_{U2}\rightarrow \frac{I(U; Y)- I(U; Z)}{I(U; Y)- I(U; V_{1}, V_{2})}$, as~$N\rightarrow \infty$. For this purpose we can follow the strategy in Section~\ref{IV-A}. So we have
\begin{eqnarray}\label{eqn21}
\lefteqn{H(S^{k}| Z^{N})}\nonumber\\
&\geq N[I(U; Y)- I(U; Z)]- H(U^{N}| S^{k}, Z^{N})
\end{eqnarray}
and for the second term in~(\ref{eqn21}) like~(\ref{eqn15}) --~(\ref{eqn17}) we have
\begin{eqnarray}\label{eqn22}
H(U^{N}| S^{k}, W, Z^{N})\leq h(P_{SB})+ P_{SB}N[I(U; Z)-\epsilon_{UZ}].
\end{eqnarray}

So, combining the above results, we have
\begin{eqnarray}\label{eqn23}
d \geq \frac{R_{U2}}{R_{U2}- \epsilon_{UY}- \epsilon_{UV_{1}V_{2}}} d_{U2}
- \frac {h(\lambda)/ N+ \lambda [I(U; V_{1}, V_{2})]+ \epsilon_{UV_{1}V_{2}}}{R_{U2}- \epsilon_{UY}- \epsilon_{UV_{1}V_{2}}}.
\end{eqnarray}

Thus we have~$d_{U2}\rightarrow \frac{I(U; Y)- I(U; Z)}{I(U; Y)- I(U; V_{1}, V_{2})}$, as~$N\rightarrow \infty$.

\section{A New Gaussian Wiretap Channel}\label{V}

\begin{figure}
\centering
\epsfig{file=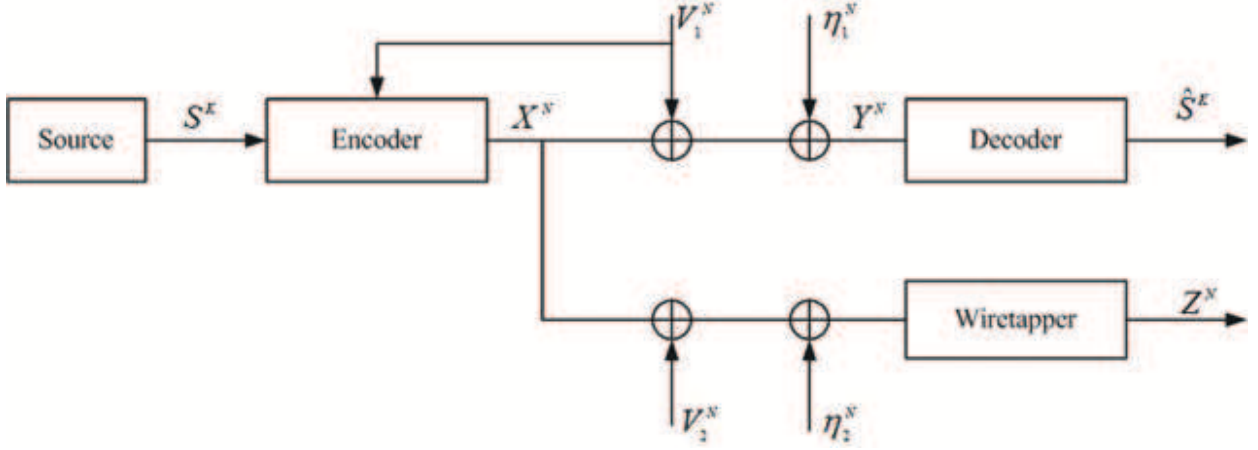,width=1\linewidth,clip=}
\caption{The new more general Gaussian wiretap channel with side information. The receiver and the wiretapper listen to the sent message from the separated channels with channel states. These channel states may have some correlation to each other.}
\label{fig:5}
\end{figure}

In this Section we extend Theorem 1 to the Gaussian case like the approach taken in~\cite{bibi23}, using the same auxiliary random variable~$U$. For the new Gaussian wiretap channel shown in Fig.~\ref{fig:5}, we have the following results based on Theorem 1.

\emph{Theorem 3:} \textit{(Theorem 1 in Gaussian case  For the Gaussian wiretap channel shown in Fig.~\ref{fig:5})} Using the auxiliary random variable~$U=X+\alpha V_{1}$, where~$\alpha$ is a real number and~$\rho_{XV_{1}}$ is the correlation coefficient of~$X$ and~$V_{1}$, we denote~$\mathcal{R}_{U}$ as the set of points~$(R, d)$ with~$R_{U1} \leq R \leq R_{U2}$,~$0\leq d\leq 1$,~$Rd=R_{U1}$, where~$R_{U1}$ and~$R_{U2}$ are defined in~(\ref{eqn7}) and~(\ref{eqn8}). By defining
\begin{eqnarray}\label{eqn24}
\mathcal{R}^{'}_{U}\triangleq \{(R^{'}, d^{'}): 0\leq R^{'}\leq R, 0\leq d^{'}\leq d, (R, d)\in \mathcal{R}_{U}\},
\end{eqnarray}
the set~$\mathcal{R}$, defined as follows, is achievable:
\begin{eqnarray}\label{eqn25}
\mathcal{R}= \bigcup_{U= X+ \alpha V_{1}, \alpha\in \mathbb{R}} \mathcal{R}^{'}_{U}.
\end{eqnarray}

\begin{proof} The proof is similar to the proof of Theorem 1. We only need to show that~$\mathcal{R}_{U}$ is achievable for the specified~$\alpha$ and~$U$. Assuming transmitter has the power constraint~$P$, the side information in the main channel satisfies~$V_{1}\sim \mathcal{N}(0, Q_{1})$, the wiretap channel has the side information, satisfying~$V_{2}\sim \mathcal{N}(0, Q_{2})$,~$\rho_{XV_{1}}$,~$\rho_{XV_{2}}$ and~$\rho_{V_{1}V_{2}}$ represent the correlation coefficient between~$X$,~$V_{1}$ and~$V_{2}$ and~$P^{'}=P[1+4\epsilon\ln 2+\frac{\rho_{XV_{1}}^{2}}{1-\rho_{XV_{1}}^{2}}]^{-1}$ (see Appendix A), we use some modification in the proof of~$\mathcal{R}_{U1}$ as follows.

 In the codebook generation, sequence~$u^{N}$ are generated according to~$f(u^{N})= \prod_{i=1}^{N} f(u_{i})$, where~$f(u_{i})\sim \mathcal{N} (0, P^{'}+ \alpha^{2}Q_{1})$ for all~$i\in \{1, 2, \ldots, N\}$.
 In the encoding process,~$x^{N}(j)=u^{N}(j)-\alpha v_{1}^{N}$.
 The intended receiver observes~$y^{N}=x^{N}+v_{1}^{N}+\eta_{1}^{N}$ and the wiretapper observes~$z^{N}=x^{N}+v_{2}^{N}+\eta_{2}^{N}$.
 As a source constraint, we should introduce potential error~$E^{X}(j)$, which represents in the encoding process and~$x^{N}(j)=u^{N}(j)-\alpha v_{1}^{N}$ does not satisfy the power constraint.

Then, provided that there is at least one sequence~$u^{N}(j)$ jointly typical with~$v_{1}^{N}$, the probability of error~$E^{X}(j)$ tends to zero. Therefore, the modifications do not influence the achievability proof of~$\mathcal{R}_{U}$. Assuming~$\epsilon$ is arbitrarily small, since~$P^{'}\rightarrow P$.
\end{proof}

Now, we calculate~$I(U; Y)$,~$I(U; V_{1},V_{2})$ and~$I(U, Z)$, with respect to~$U=X+\alpha V_{1}$. We have
\begin{eqnarray}\label{eqn26}
\lefteqn{I(U; Y)}\nonumber\\
&\!\!\!\!\!\!\!\!\!\!\!\!\!\!\!\!\!\!\!\!\!\!\!\!\!\!\!\!\!\!\!\!\!\!\!\!\!\!\!\!\!
=\frac{1}{2}\log[\frac{(P+\alpha^{2}Q_{1}+2\alpha \rho_{X V_{1}}\sqrt{PQ_{1}})(P+ Q_{1}+ N_{1} +2\rho_{X V_{1}}\sqrt{P Q_{1}})}{(P+\alpha^{2}Q_{1}+2\alpha \rho_{X V_{1}}\sqrt{PQ_{1}})(P+ Q_{1}+ N_{1}+2\rho_{X V_{1}}\sqrt{P Q_{1}})-(P+ \alpha Q_{1}+(\alpha+1)\rho_{X V_{1}}\sqrt{P Q_{1}})^{2}}]\\
&\!\!\!\!\!\!\!\!\!\!\!\!\!\!\!\!\!\!\!\!\!\!\!\!\!\!\!\!\!\!\!\!\!\!\!\!\!\!\!\!\!\!\!\!\!\!\!\!\!\!\!\!\!\!\!\!\!\!\!\!
\!\!\!\!\!\!\!\!\!\!\!\!\!\!\!\!\!\!\!\!\!\!\!\!\!\!\!\!\!\!\!\!\!\!\!\!\!\!\!\!\!\!\!\!\!\!\!\!
I(U; V_{1}, V_{2})= \frac{1}{2}\log[\frac{(1-\rho_{V_{1}V_{2}}^{2})(P+\alpha^{2}Q_{1}+2\alpha \rho_{X V_{1}}\sqrt{PQ_{1}})}{P(1-\rho_{XV_{1}}^{2}-\rho_{XV_{2}}^{2}-\rho_{V_{1}V_{2}}^{2}+2\rho_{XV_{1}}\rho_{XV_{2}}\rho_{V_{1}V_{2}})}]\\
&\!\!\!\!\!\!\!\!\!\!\!\!\!\!\!\!\!\!\!\!\!\!\!\!\!\!\!\!\!\!\!\!\!\!\!\!\!\!\!\!\!\!\!\!\!\!\!\!\!\!\!\!\!\!\!\!\!\!
\!\!\!\!\!\!\!\!\!\!\!\!\!\!\!\!\!\!\!\!\!\!\!\!\!\!\!\!\!\!\!\!\!\!\!\!\!\!\!\!\!\!\!\!\!\!\!\!\!\!\!\!\!\!\!\!\!\!\!\!\!\!
\!\!\!\!\!\!\!\!\!\!\!\!\!\!\!\!\!\!\!\!\!\!\!\!\!\!\!\!\!\!\!\!\!\!\!\!\!\!\!\!\!\!\!\!\!\!\!\!\!\!\!\!\!\!\!\!\!\!\!\!\!\!
\!\!\!\!\!\!\!\!\!\!\!\!\!\!\!\!\!\!\!\!\!\!\!\!\!\!\!\!\!\!\!\!\!\!\!\!\!\!\!\!\!\!\!\!\!\!\!\!\!\!\!\!\!\!\!\!\!\!\!\!
I(U; Z)\nonumber\\
&= \frac{1}{2}\log[\frac{(P+\alpha^{2}Q_{1}+2\alpha \rho_{X V_{1}}\sqrt{PQ_{1}})(P+ Q_{2}+ N_{2} +2\rho_{X V_{2}}\sqrt{P Q_{2}})}{(P+\alpha^{2}Q_{1}+2\alpha \rho_{X V_{1}}\sqrt{PQ_{1}})(P+ Q_{2}+ N_{2}+2\rho_{X V_{2}}\sqrt{P Q_{2}})-(P+ \rho_{X V_{2}}\sqrt{P Q_{2}}+ \alpha \rho_{X V_{1}}\sqrt{PQ_{1}}+\alpha\rho_{V_{1}V_{2}}\sqrt{Q_{1}Q_{2}})^{2}}]\nonumber\\
\end{eqnarray}

Then, we introduce \textit{Leakage Function}~$\Delta I(\alpha)$ which is defined as~$\Delta I(\alpha)= I(U; Z)- I(U; V_{1}V_{2})$. Thus, we have
\begin{eqnarray}\label{eqn29}
\lefteqn{\Delta I(\alpha)= I(U; Z)- I(U; V_{1}V_{2})}\nonumber\\
&=\frac{1}{2}\log[\frac{P(P+ Q_{2}+ N_{2} +2\rho_{X V_{2}}\sqrt{P Q_{2}})(1-\rho_{XV_{1}}^{2}-\rho_{XV_{2}}^{2}-\rho_{V_{1}V_{2}}^{2}+2\rho_{XV_{1}}\rho_{XV_{2}}\rho_{V_{1}V_{2}})}{(P+\alpha^{2}Q_{1}+2\alpha \rho_{X V_{1}}\sqrt{PQ_{1}})(P+ Q_{2}+ N_{2}+2\rho_{X V_{2}}\sqrt{P Q_{2}})-(P+ \rho_{X V_{2}}\sqrt{P Q_{2}}+ \alpha \rho_{X V_{1}}\sqrt{PQ_{1}}+\alpha\rho_{V_{1}V_{2}}\sqrt{Q_{1}Q_{2}})^{2}}]
\end{eqnarray}

Hence
\begin{eqnarray}\label{eqn30}
\Delta I(0)=\frac{1}{2}\log[\frac{(P+ Q_{2}+ N_{2} +2\rho_{X V_{2}}\sqrt{P Q_{2}})(1-\rho_{XV_{1}}^{2}-\rho_{XV_{2}}^{2}-\rho_{V_{1}V_{2}}^{2}+2\rho_{XV_{1}}\rho_{XV_{2}}\rho_{V_{1}V_{2}})}{Q_{2}(1- \rho_{XV_{2}}^{2})+ N_{2}+ 2\rho_{X V_{2}}\sqrt{P Q_{2}}}]> 0
\end{eqnarray}
and we can find two points~$\alpha_{0}$ and~$\alpha_{-0}$ in which
\begin{eqnarray}\label{eqn31}
\lefteqn{\!\!\!\!\!\!\!\!\!\Delta I(\alpha_{0})= \Delta I(\alpha_{-0})=0.}
\end{eqnarray}

Furthermore, there is a point~$\alpha^{*}$ in which~$\Delta I(\alpha)$ is maximized, i.e.,
\begin{eqnarray}\label{eqn32}
\lefteqn{\!\!\!\!\!\!\!\!\!\!\!\!\!\!\!\!\!\!\!\!\!\!\!\!\!\!\!\!\!\!\!\!\!\!\!\!\!\!\!\!\!\!\!\!\!\!\!\!\!\!\!\!\!\!\!\!\!\!\!\!\!\!\!\!\!\!\!\!\!
\!\!\!\!\!\!\!\!\!\!\!\!\!\!\!\!\!\!\!\!\!\!\!\!\!\!\!\!\!\!\!\!\!\!\!\!\!\!\!\!\!\!\!\!\!\!\!\!\!\!\!\!\!\!\!\!\!\!\!\!\!\!\!\!
\alpha^{*}=-\frac{(\rho_{X V_{1}}\sqrt{PQ_{1}}+\rho_{V_{1}V_{2}}\sqrt{Q_{1}Q_{2}})(P+\rho_{X V_{2}}\sqrt{PQ_{1}})-
\rho_{XV_{1}}\sqrt{PQ_{1}}(P+N_{2}+Q_{2}+2\rho_{XV_{2}}\sqrt{PQ_{2}})}{(\rho_{X V_{1}}\sqrt{PQ_{1}}+\rho_{V_{1}V_{2}}
\sqrt{Q_{1}Q_{2}})^2-2Q_{1}(P+N_{2}+Q_{2}+2\rho_{XV_{2}}\sqrt{PQ_{2}})}}
\end{eqnarray}
where
\begin{eqnarray}\label{eqn33}
\max\Delta I(\alpha)=\Delta I(\alpha^{*})
\end{eqnarray}

Now, we want to study the leakage function. So, denote~$R(\alpha)=I(U; Y)- I(U; V_{1}, V_{2})$ and~$R_{Z}(\alpha)=I(U; Y)- I(U; Z)$. Because of the complexity of the results, we consider two special cases.

\subsection{Case I}
As the first condition, we assume that~$Q_{1}=Q_{2}=Q$,~$\rho_{XV_{1}}=\rho_{XV_{2}}=0$ and~$\rho_{V_{1}V_{2}}=1$. In this case our model reduces to the channel introduced~\cite{bibi23} and we have
\begin{eqnarray}\label{eqn34}
\lefteqn{\!\!\!\!\!\!\!\!\!\!\!\!\!\!\!\!\!\!\!\!\!\!\!\!\!\!\!\!\!\!\!\!\!\!\!\!\!\!\!\!\!\!\!\!\!\!\!\!\!\!
R(\alpha)=\frac{1}{2}\log[\frac{P(P+Q+N_{1})}{(P+\alpha^{2}Q)(P+Q+N_{1})-(P+\alpha Q)^{2}}]}
\end{eqnarray}
which is maximized by~$\alpha^{*}=\frac{P}{P+N}$ as described in~\cite{bibi24} and achieves~$C_{M}=\frac{1}{2}\log[\frac{P+N}{N}]$, in which~$C_{M}$ is the maximum rate of the main channel. It can be found easily that~$R(\alpha)$ is an increasing function with respect to~$\alpha$ as~$\alpha<\alpha^{*}$, a decreasing function with respect to~$\alpha$ as~$\alpha>\alpha^{*}$.

Similarly, the rate~$R_{Z}$ has two extremum points in~$\alpha=1$ and~$\alpha=-\frac{P}{Q}$ and it can be shown that~$R_{Z}$ is a decreasing function with respect to~$\alpha$ as~$\alpha<-\frac{P}{Q}$ or~$1<\alpha$ and an increasing function with respect to ~$\alpha$ as~$-\frac{P}{Q}<\alpha<1$.
Then we state the following result.

\emph{Theorem 4:} For the new Gaussian wiretap channel with side information illustrated in Fig. 5, under the special conditions explained in Case I, rate equivocation pair~$(R, d)$ is achievable if
\begin{eqnarray}\label{eqn35}
\lefteqn{\!\!\!\!\!\!\!\!\!\!\!\!\!\!\!\!\!\!\!\!\!\!\!\!\!\!\!\!\!\!\!\!\!\!\!\!\!\!\!\!\!\!\!\!\!\!\!\!\!\!\!\!\!\!\!\!\!\!\!\!
\!\!\!\!\!\!\!\!\!\!\!\!\!\!\!\!\!\!\!\!\!\!\!\!\!\!\!\!\!\!\!\!\!\!\!\!\!\!\!\!
\!\!\!\!\!\!\!\!\!\!\!\!\!\!\!\!\!\!\!\!\!\!\!\!\!\!\!\!\!\!\!\!\!\!\!\!\!\!\!\!\!\!\!\!\!\!\!\!
R\leq C_{M}}\nonumber\\
\lefteqn{\!\!\!\!\!\!\!\!\!\!\!\!\!\!\!\!\!\!\!\!\!\!\!\!\!\!\!\!\!\!\!\!\!\!\!\!\!\!\!\!\!\!\!\!\!\!\!\!\!\!\!\!\!\!\!\!\!\!\!\!
\!\!\!\!\!\!\!\!\!\!\!\!\!\!\!\!\!\!\!\!\!\!\!\!\!\!\!\!\!\!\!\!\!\!\!\!\!\!\!\!
\!\!\!\!\!\!\!\!\!\!\!\!\!\!\!\!\!\!\!\!\!\!\!\!\!\!\!\!\!\!\!\!\!\!\!\!\!\!\!\!\!\!\!\!\!\!\!\!
d\leq 1}\nonumber\\
Rd \leq \left\{
   \begin{array}{ll}
        C_{M} & 0 < P \leq P_{1}\\
        \left\{\begin{array}{ll}
            R(\alpha_{0}) & R\leq R(\alpha_{0})\\
            R_{Z}(\alpha) & R(\alpha_{0})\leq R\leq C_{M}\\
        \end{array}\right.
        & P_{1}\leq P\leq P_{2}\\
        \left\{\begin{array}{ll}
            R_{Z}(1) & R\leq R(1)\\
            R_{Z}(\alpha) & R(1)\leq R \leq C_{M}\\
        \end{array}\right.
        & P_{2}\leq P\\
    \end{array}\right.
\end{eqnarray}
where
\begin{eqnarray}\label{eqn36}
\lefteqn{\!\!\!\!\!\!\!\!\!\!\!\!\!\!\!\!\!\!\!\!\!\!\!\!\!\!\!\!\!\!\!\!\!\!\!\!\!\!\!\!\!\!\!\!\!\!\!\!\!\!\!\!\!\!\!\!\!\!\!\!\!
\!\!\!\!\!\!\!\!\!\!\!\!\!\!\!\!\!\!\!\!\!\!\!\!\!\!\!\!\!
P_{1}=-N_{1}- \frac{Q}{2}+ \frac{\sqrt{Q^{2}+4QN_{2}}}{2}}\\
P_{2}=- \frac{Q}{2}+ \frac{\sqrt{Q^{2}+4Q(N_{1}+ N_{2})}}{2}
\end{eqnarray}

We should note that this rate equivocation pair is similar the one presented in~\cite{bibi23} and the proof can be found there. It is clear that under the assumed conditions, our channel reduces to the previous model~\cite{bibi23} and we obtain similar result.

 \emph{Corollary 1:} ~\cite[Theorem 4-5]{bibi23} For the proposed Gaussian wiretap channel with side information in Case I, the side information helps to achieve larger rate equivocation region. The proof is similar to the one in~\cite{bibi23}.

\subsection{Case II}

As the second special case, we assume that~$Q_{1}=Q_{2}=Q$,~$\rho_{XV_{1}}=\rho_{XV_{2}}=\rho_{V_{1}V_{2}}=0$ and~$N_{1}\neq N_{2}$. In this case we have

\begin{eqnarray}\label{eqn38}
\lefteqn{\!\!\!\!\!\!\!\!\!\!\!\!\!\!\!\!\!\!\!\!\!\!\!\!\!\!\!\!\!\!\!\!\!\!\!\!\!\!\!\!\!\!\!\!\!\!\!\!\!\!
R(\alpha)=\frac{1}{2}\log[\frac{P(P+Q+N_{1})}{(P+\alpha^{2}Q)(P+Q+N_{1})-(P+\alpha Q)^{2}}]}
\end{eqnarray}
which is maximized by~$\alpha^{*}=\frac{P}{P+N}$ and achieves~$C_{M}$. It can be found that the rate~$R_{Z}$ has two extremum points in~$\alpha=1$ and~$\alpha= -\frac{P}{Q}$. As we can see, this points are similar to the one for the functions in the previous case. So, we state the following result for this case.

\emph{Theorem 5:} For the proposed Gaussian wiretap channel with side information in Fig. 5, under the special conditions, a rate pair~$(R, d)$ is achievable if

\begin{eqnarray}\label{eqn39}
\lefteqn{\!\!\!\!\!\!\!\!\!\!\!\!\!\!\!\!\!\!\!\!\!\!\!\!\!\!\!\!\!\!\!\!\!\!\!\!\!\!\!\!\!\!\!
\!\!\!\!\!\!\!\!\!\!\!\!\!\!\!\!\!\!\!\!\!\!\!\!\!\!
\!\!\!\!\!\!\!\!\!\!\!\!\!\!\!\!\!\!\!\!\!\!\!\!\!\!\!\!\!\!\!
\!\!\!\!\!\!\!\!\!\!\!\!\!\!\!\!\!\!\!\!\!\!\!\!\!\!\!\!\!\!\!\!\!\!\!\!\!\!\!\!\!\!\!\!\!\!\!\!
R\leq C_{M}}\nonumber\\
\lefteqn{\!\!\!\!\!\!\!\!\!\!\!\!\!\!\!\!\!\!\!\!\!\!\!\!\!\!\!\!\!\!\!\!\!\!\!\!\!\!\!\!\!\!\!
\!\!\!\!\!\!\!\!\!\!\!\!\!\!\!\!\!\!\!\!\!\!
\!\!\!\!\!\!\!\!\!\!\!\!\!\!\!\!\!\!\!\!\!\!\!\!\!\!\!\!\!\!\!\!\!\!\!
\!\!\!\!\!\!\!\!\!\!\!\!\!\!\!\!\!\!\!\!\!\!\!\!\!\!\!\!\!\!\!\!\!\!\!\!\!\!\!\!\!\!\!\!\!\!\!\!
d\leq 1}\nonumber\\
Rd \leq \left\{
   \begin{array}{ll}
        C_{M} & 0 < P \leq P_{3}\\
        \left\{\begin{array}{ll}
            R(\alpha_{0}) & R\leq R(\alpha_{0})\\
            R_{Z}(\alpha) & R(\alpha_{0})\leq R\leq C_{M}\\
        \end{array}\right.
        & P_{3}\leq P\leq P_{4}\\
        \left\{\begin{array}{ll}
            R_{Z}(1) & R\leq R(1)\\
            R_{Z}(\alpha) & R(1)\leq R \leq C_{M}\\
        \end{array}\right.
        & P_{4}\leq P\\
    \end{array}\right.
\end{eqnarray}
where
\begin{eqnarray}\label{eqn40}
\lefteqn{\!\!\!\!\!\!\!\!\!\!\!\!\!\!\!\!\!\!\!\!\!\!\!\!\!\!\!\!\!\!\!\!\!\!\!\!\!\!\!\!\!\!\!\!\!\!\!\!\!\!\!\!
\!\!\!\!\!\!\!\!\!\!\!\!\!\!\!\!\!\!\!\!\!\!\!\!\!\!\!\!\!\!\!\!\!\!\!\!\!\!
P_{3}=\frac{(Q-2N_{1})+ \sqrt{5Q^{2}+4Q(N_{2}-N_{1})}}{2}}\\
P_{4}=\frac{Q}{2}+ \frac{\sqrt{5Q^{2}+4Q N_{2}}}{2}
\end{eqnarray}

So the obtained results are similar to the previous case for Gaussian wiretap channel with side information.

\section*{Conclusion}\label{conc}
In this paper, a new applicable wiretap channel with side information was introduced. In this channel, the previous models were generalized.
An achievable equivocation rate region for this channel was derived and then, our result were extended to the Gaussian case.

\bibliography{Wiretap Channel}
\bibliographystyle{IEEE}

\appendices
\section{The Condition on Average Power Constraint}

In this Section we apply the following lemma to the condition on the average power constraint by letting~$P^{'}=P[1+4\epsilon\ln 2+\frac{\rho_{XV_{1}}^{2}}{1-\rho_{XV_{1}}^{2}}]^{-1}$.

\emph{Lemma 1:} Assume that~$X^{N}$ and~$V_{1}^{N}$ are two sequences of i.i.d random variables~$X\sim \mathcal{N}(0, \sigma_{X})$ and~$V_{1}\sim \mathcal{N}(0, \sigma_{V_{1}})$, respectively, with correlation coefficient~$\rho_{XV_{1}}$. Let~$U^{N}=X^{N}+\alpha V_{1}^{N}$, where~$\alpha$ is a constant real number. If~$(u^{N}, v^{N})\in T^{N}_{U, V_{1}}(\epsilon)$, for any~$\epsilon> 0$, and~$\sigma_{X}\leq P[1+4\epsilon\ln 2+\frac{\rho_{XV_{1}}^{2}}{1-\rho_{XV_{1}}^{2}}]^{-1}$,
then~$[\sum^{N}_{i=1}x^{2}_{i}]/N\leq P$.
\begin{proof}
Since~$X^{N}$ and~$V^{N}_{1}$ are two sequences of i.i.d Gaussian random variables then~$U^{N}\sim \mathcal{N}(0, \sigma_{X}+\alpha^{2}\sigma_{V_{1}}+2\alpha\sigma_{X}\sigma_{V_{1}})$. Furthermore~$(u^{N}, v^{N})\in T^{N}_{U, V_{1}}(\epsilon)$ implies that

\begin{eqnarray}\label{eqn26}
\lefteqn{\epsilon > |-\frac{1}{N}\log p(u^{N}, v^{N})- H(U, V)\mid}\nonumber\\
&\!\!\!\!\!\!\!\!\!\!\!\!\!\!\!\!\!\!\!\!\!\!\!\!\!\!\!\!\!\!\!\!\!\!\!\!\!\!\!\!\!\!\!\!\!\!\!\!\!\!\!\!\!\!\!\!\!\!\!\!\!\!\!\!\!\!\!
\!\!\!\!\!\!\!\!\!\!\!\!\!\!\!\!\!\!\!\!\!\!\!\!\!\!\!\!\!\!\!\!\!
=\mid-\frac{1}{N}\log p(u^{N}, v^{N})- H(V)-H(U| V)|\nonumber\\
&\!\!\!\!\!\!\!\!\!\!\!\!\!\!\!\!\!\!\!\!\!\!\!\!\!\!\!\!\!\!\!\!\!\!\!\!\!\!\!\!\!\!\!\!\!\!\!\!\!\!\!\!\!\!\!\!\!\!\!\!\!\!\!\!\!\!\!
\!\!\!\!\!\!\!\!\!\!\!\!\!\!\!\!\!\!\!\!\!\!\!\!\!\!\!\!\!\! =|-\frac{1}{N}\log p(u^{N}, v^{N})- H(V)-H(X| V)|\nonumber\\
&\!\!\!\!\!\!\!\!\!\!\!\!\!\!\!\!\!\!\!\!\!\!\!\!\!\!\!\!\!\!\!\!\!\!\!\!\!\!\!\!\!\!\!\!\!\!\!\!\!\!\!\!\!\!\!\!\!\!\!\!\!\!\!\!\!\!\!
\!\!\!\!\!\!\!\!\!\!\!\!\!\!\!\!\!\!\!\!\!\!\!\!\!\!\!\!\!\!\!\!\!\!\!\!\!\!\!\!\!\!\!\!\!\!\!\! \!\!\!\!\!\!\!\!\!\!\!\!\!\!\!\!\!\!\!\!\!\!\!\!\!\!\!\!\!\!\!\!\!\!\!\!\! 2 \epsilon > |-\frac{1}{N}\log p(u^{N}, v^{N})+\frac{1}{N}\log p(v^{N})- H(X| V)|\nonumber\\
&\!\!\!\!\!\!\!\!\!\!\!\!\!\!\!\!\!\!\!\!\!\!\!\!\!\!\!\!\!\!\!\!\!\!\!\!\!\!\!\!\!\!\!\!\!\!\!\!\!\!\!\!\!\!\!\!\!\!\!\!\!\!\!\!\!\!
\!\!\!\!\!\!\!\!\!\!\!\!\!\!\!\!\!\!\!\!\!\!\!\!\!\!\!\!\!\!\!\!\!\!\!\!\!\!\!\!\!\!\!\!\!\!\!\!\!\!\!\!\! = |-\frac{1}{N}\log p(u^{N}| v^{N})- H(X| V)|\nonumber\\
&\!\!\!\!\!\!\!\!\!\!\!\!\!\!\!\!\!\!\!\!\!\!\!\!\!\!\!\!\!\!\!\!\!\!\!\!\!\!\!\!\!\!\!\!\!\!\!\!\!\!\!\!\!\!\!\!\!\!\!\!\!\!\!\!\!\!
\!\!\!\!\!\!\!\!\!\!\!\!\!\!\!\!\!\!\!\!\!\!\!\!\!\!\!\!\!\!\!\!\!\!\!\!\!\!\!\!\!\!\!\!\!\!\!\!\!\!\!\!\! = |-\frac{1}{N}\log p(x^{N}| v^{N})- H(X| V)|\nonumber\\
&\!\!\!\!\!\!\!\!\!\!\!\!\!\!\!\!\!\!\!\!\!\!\!\!\!\!\!\!\!\!\!\!\!\!\!\!\!\!\!\!\!\!\!\!\!\!\!\!\!\!\!\!\!\!\!\!\!\!\!\!\!\!\!\!\!\!
\!\!\!\!\!\!\!\!\!\!\!\!\!\!\!\!\!\!\!\!\!\!\!\!\!\!\!\!\!\!\!\!\!\!\!\!\!\!\! =|-\frac{1}{N}\sum^{N}_{i=1}\log p(x_{i}| v_{1, i})- H(X| V)|\nonumber\\
&\quad\quad\quad\quad\quad\quad\!\!=^{(d)} \frac{1}{\ln 2}|\frac{1}{2N(1-\rho_{XV_{1}}^{2})}\sum^{N}_{i=1}{\frac{x_{i}^{2}}{\sigma_{X}^{2}}-2\rho_{X V_{1}}\frac{x_{i}v_{1, i}}{\sigma_{X}\sigma_{V_{1}}} +\rho_{XV_{1}}^{2}\frac{v_{1i}^{2}}{\sigma_{V_{1}^{2}}}+\frac{1}{2}ln [2\pi \sigma_{X}^{2}(1- \rho_{XV_{1}}^{2})]}- H(X| V)|\nonumber\\
&\quad\quad\quad\!\!=^{(e)} \frac{1}{\ln 2}|\frac{\sum^{N}_{i=1} x_{i}^{2}}{2N(1-\rho_{XV_{1}}^{2})\sigma_{X}^{2}}-\frac{\rho_{X V_{1}}^{2}}{2(1- \rho_{XV_{1}}^{2})}+\frac{1}{2}\ln [2\pi \sigma_{X}^{2}(1- \rho_{XV_{1}}^{2})]-\frac{1}{2}\ln [2\pi e \sigma_{X}^{2}(1- \rho_{XV_{1}}^{2})]|\nonumber\\
&\!\!\!\!\!\!\!\!\!\!\!\!\!\!\!\!\!\!\!\!\!\!\!\!\!\!\!\!\!\!\!\!\!\!\!\!\!\!\!\!\!\!\!\!\!\!\!\!\!\!\!\!\!\!\!\!\!\!\!\!\!\!\!\!\!\!
\!\!\!\!\!\!\!\!\!\!\!\!\!\!\!\!\!\!\!\!\!\!\!\!\!\!\!\!\!\!\!\!\!\!\!\!\!\!\!\!\!\! =\frac{1}{\ln 2}|\frac{\sum^{N}_{i=1} x_{i}^{2}}{2N(1-\rho_{XV_{1}}^{2})\sigma_{X}^{2}}-\frac{\rho_{X V_{1}}^{2}}{2(1- \rho_{XV_{1}}^{2})}- \frac{1}{2}|\nonumber\\
\end{eqnarray}
where~$(d)$ is because of the jointly distribution function of~$(x_{i}, v_{1, i})$ and~$(e)$ is because that
\begin{eqnarray}\label{eqn27}
\lefteqn{\!\!\!\!\!\!\!\!\!\!\!\!\!\!\!\!\!\!\!\!\!\!\!\!\!\!H(X| V)=H(X)-I(X; V_{1})=\frac{1}{2}\ln(2\pi e \sigma_{X}^{2})- \frac{1}{2}\ln (\frac{\sigma^{2}_{X}\sigma^{2}_{V_{1}}}{\sigma^{2}_{X}\sigma^{2}_{V_{1}}- \sigma^{2}_{XV_{1}}})}\nonumber\\
&\!\!\!\!\!\!\!\! =\frac{1}{2}\ln [2\pi e \sigma_{X}^{2}(1- \sigma^{2}_{XV_{1}})]\nonumber\\
\end{eqnarray}
Thus
\begin{eqnarray}\label{eqn28}
\frac{1}{N}\sum^{N}_{i=1} x^{2}_{i}< \sigma_{X}^{2}[ 4 \epsilon \ln 2+ \frac{\rho_{XV_{1}}^{2}}{1- \rho_{XV_{1}}^{2}}+1]
\end{eqnarray}
and with the condition on the average power constraint~$\sigma_{X}\leq P[1+4\epsilon\ln 2+\frac{\rho_{XV_{1}}^{2}}{1-\rho_{XV_{1}}^{2}}]^{-1}$, we have~$[\sum^{N}_{i=1}x^{2}_{i}]/N\leq P$.
\end{proof}
\end{document}